\documentclass[12pt]{iopart}

\usepackage{amsfonts,amssymb}
\usepackage{graphicx,epsfig}
\usepackage{bm}

\def\be{\begin{equation}}
\def\ee{\end{equation}}
\def\bea{\begin{eqnarray}}
\def\eea{\end{eqnarray}}
\def\bfa{\begin{mathletters}}
\def\ema{\end{mathletters}}

\def\0{\overline{0}}

\def\q0{\underline{0}}

\def\one{\leavevmode\hbox{\small1\normalsize\kern-.33em1}}

\begin{document}

\title{Off-diagonal correlations in a one-dimensional gas of dipolar bosons}
\author{Tommaso Roscilde$^1$ and Massimo Boninsegni$^2$}
\address{$^1$ Laboratoire de Physique, Ecole Normale Sup\'erieure de Lyon, 46 All\'ee d'Italie,
69007 Lyon, France}
\address{$^2$ Department of Physics and Astronomy, University of Alberta, 248L CEB, 11322-89 Ave,
Edmonton, Alberta T6G 2G7, Canada}

\begin{abstract}
We present  a quantum Monte Carlo study of the one-body density matrix (OBDM) and the momentum 
distribution of one-dimensional dipolar bosons, with dipole moments polarized perpendicular to the 
direction of confinement. We observe that the long-range nature of the dipole interaction has dramatic 
effects on the off-diagonal correlations: although the dipoles never crystallize, the system goes from 
a quasi-condensate regime at low interactions to a regime in which quasi-condensation is 
discarded, in favor of quasi-solidity.
For all strengths of the dipolar interaction, the OBDM shows an oscillatory behavior coexisting with an 
overall algebraic decay; and the momentum distribution shows sharp kinks at the wavevectors
of the oscillations, $Q = \pm 2\pi n$ (where $n$ is the atom density), beyond which it is strongly
suppressed. This \emph{momentum filtering} effect introduces a characteristic scale in the 
momentum distribution, which can be arbitrarily squeezed by lowering the atom density.
This shows that one-dimensional dipolar Bose gases, realized e.g. by trapped dipolar
molecules, show strong signatures of the dipolar interaction in time-of-flight measurements. 
\end{abstract}

\section{Introduction}

 Trapped quantum degenerate gases are currently offering the possibility of 
 a thorough investigation of fundamental models of correlated quantum
 liquids, spanning a wide range of regimes from weak to strong
 interactions \cite{Blochetal08}.  A particularly appealing aspect of 
 cold atoms is indeed the tunability of the interparticle interactions. 
 So far most of the experiments have been performed in a regime 
 in which the interaction can be faithfully described as a contact
 $s$-wave interaction; tuning an applied magnetic field close to a 
 Feshbach resonance allows to control continuously the magnitude
 and sign of the associated $s$-wave scattering length  
 \cite{Chinetal09}. The use of Feshbach resonances allows
 in particular to suppress almost completely the contact interaction, 
 leading to the observation of the effects of weaker magnetic dipole-dipole 
 interactions for atoms which have a net magnetic moment, such as 
 $^{52}$Cr \cite{Lahayeetal07, Kochetal08, Lahayeetal08}.  
  On the other hand, the recent realization of a stable degenerate gas
of heteronuclear molecules \cite{Nietal08} appears to pave the way to the exploration 
of the physics of quantum fluids with strong (and variable) 
electric dipole-dipole interactions; in fact the application of an electric field
on the heteronuclear molecules, or the dressing of molecules with a microwave 
field, can induce a tunable electric dipole. 

  On the theoretical side, the presence of the dipolar interaction has been 
 shown to lead to many intriguing effects when the system is confined to 
 two spatial dimensions, such as roton-like excitations
 in the liquid regime \cite{Lahayeetal09}, spontaneous crystallization \cite{Buechleretal07}, 
 and even supersolidity in presence of an optical lattice \cite{Lahayeetal09}. 
 In the case of dipolar bosons confined to one spatial dimension, in the
 absence of an optical lattice 
 Mermin-Wagner-Hohenberg theorem forbids crystallization even at $T=0$
 due to unboundedly strong quantum fluctuations.
 Hence the system can be described as a Luttinger liquid for all strengths of the 
 dipolar interaction \cite{Citroetal07, Citroetal08, Pedrietal08, DePaloetal08}.
 While it can develop arbitrarily strong diagonal correlations, resulting
 in a so-called super-Tonks behavior
\cite{Arkhipovetal05, AstrakharchikL08}, such correlations are always decaying 
as a power of the distance. 

 All the existing studies of the one-dimensional dipolar Bose gas have mainly 
 focused on diagonal correlations and on the static structure factor \cite{Citroetal07, Citroetal08, DePaloetal08,
 Arkhipovetal05, AstrakharchikL08}. Yet the most
 natural observables in cold atom experiments are off-diagonal correlations, 
 namely the one-body density matrix (OBDM) and its Fourier transform, the momentum distribution,
 which is obtained from time-of-flight measurements. 
 In this paper we focus our attention on these quantities, which we compute numerically exactly 
 by quantum Monte Carlo simulations. 
 In particular, we show that off-diagonal correlations exhibit very strong signatures
 of dipolar physics. In fact the OBDM has a characteristic oscillatory behavior at
 the ordering vector $Q$ of the Wigner crystal obtained in the classical limit. Such a behavior 
 translates into a strong feature in the momentum distribution, namely a sharp kink
 at the $Q$ vector, and a strong suppression of the momentum population for 
 larger wavevectors. This effect of momentum filtering in the range $[-Q,Q]$ is 
 completely controlled by the density of the system, so that the momentum distribution
 can be arbitrarily squeezed in momentum space, although the system \emph{never} 
 attains condensation at finite density. A study of the two-body problem suggests that 
 momentum filtering is a special feature of long-range interactions already at the 
 few-body level, although its manifestation in the $N$-body problem is much more pronounced. 
 
  The paper is structured as follows. Section \ref{sec:model} describes the system
  Hamiltonian and the numerical method; Section \ref{sec:twobody} presents a study
  of off-diagonal correlations in the two-body problem with dipolar 
  interactions, contrasting it with the case of contact interactions; Section \ref{sec:corrfunc}
  focuses on the OBDM as compared to predictions from Luttinger liquid theory; 
  and finally Section \ref{sec:nk} discusses the results concerning the
  momentum distribution.

\section{\label{sec:model} Model and methods} 
  
  We investigate a system of $N$ bosons confined to a one dimensional box of length $L$
  with periodic boundary conditions. The system Hamiltonian reads 
  \begin{equation}
  {\cal H} =  -\frac{\hbar^2}{2m} \sum_{i=1}^{N} \frac{d^2}{dx_i^2} +  \sum_{i<j} V(x_i - x_j) ~.
  \label{e.Ham}
   \end{equation}
  In the following, otherwise specified, we will focus on the dipolar potential
 \begin{equation}
 V_{\rm dip}(x-x') = \frac{V_d}{|x-x'|^3} 
 \label{e.pot}
 \end{equation}
Introducing the particle density $n = N/L$ and the rescaled variables $\tilde{x_i} = x_i/r_0$
(where $r_0 = mV_d/\hbar^2$ is an effective length associated with the potential), 
the Hamiltonian with dipolar interactions can be rewritten in the convenient form
\begin{equation}
  {\cal H}({\rm Ryd}) =  (nr_0)^2 \left[ -\frac{1}{2} \sum_{i=1}^{N} \frac{d^2}{d\tilde{x}_i^2} + (nr_0) \sum_{i<j} \frac{1}{|\tilde{x}_i - \tilde{x}_j|^3} \right]~.
  \label{e.Ham2}
\end{equation}
Here the Hamiltonian is expressed in effective Rydberg units, Ryd $= \hbar^2/(mr_0^2)$
\cite{note-comparison}.
Hence $nr_0$ gives the dimensionless strength of the dipolar potential, and it contains
a fundamental property of scale invariance, typical of power-law decaying potentials:
rescaling the density $n \to n/\lambda$ and the effective length $r_0 \to \lambda r_0$, 
the physics of the system is unchanged (apart from a rescaling of the box length, $L\to \lambda L$,
and, accordingly, of the  positions,  $x \to \lambda x$).
 In the following we will compare the behavior of the dipolar system with that of the 
 Lieb-Liniger gas, characterized by the $\delta$-potential
 \begin{equation}
 V_{\delta}(x-x') = g~ \delta(x-x')~.
 \end{equation} 
 with repulsive nature, $g>0$. The associated scattering length is given by $a = - 2\hbar^2/(mg)$. 
 The dimensionless strength of the potential can be measured by $(n|a|)^{-1}$. 

We investigate the ground-state properties of the 1D gas of dipolar bosons by 
quantum Monte Carlo simulations, based on the continuous-space Worm Algorithm, which
allows one to obtain the one-body density matrix straightforwardly \cite{Boninsegnietal06,Boninsegni07}.
While the method is strictly valid for finite temperatures only, the zero temperature 
 behavior can be recovered for low enough temperatures.
 In addition to Monte Carlo simulations,  we also use standard exact diagonalization
 to study the problem of two interacting bosons, whose results are presented in the
 next section.    
 
\section{\label{sec:twobody} Two-body problem}

 The study of the off-diagonal correlations of the two body problem turns out to be
 quite insightful in the perspective of the $N$-body problem, as some fundamental
 traits of the latter are already exhibited by the former. We numerically evaluate the
 ground-state wavefunction $\phi(r)$ for the relative coordinate of the two-body problem
 on a $L=2 l$ ring (where $l$ is an arbitrary length unit), and the associated momentum distribution
\begin{equation}  
n(k) = \frac{2}{L} \left| \int dr ~e^{ikr}~ \phi(r) \right|^2
\end{equation}
defined on discrete momenta $k = 2\pi p/L$, $p\in Z$.
 
\begin{figure}[h!]
\begin{center}
\includegraphics[width=0.75\linewidth]{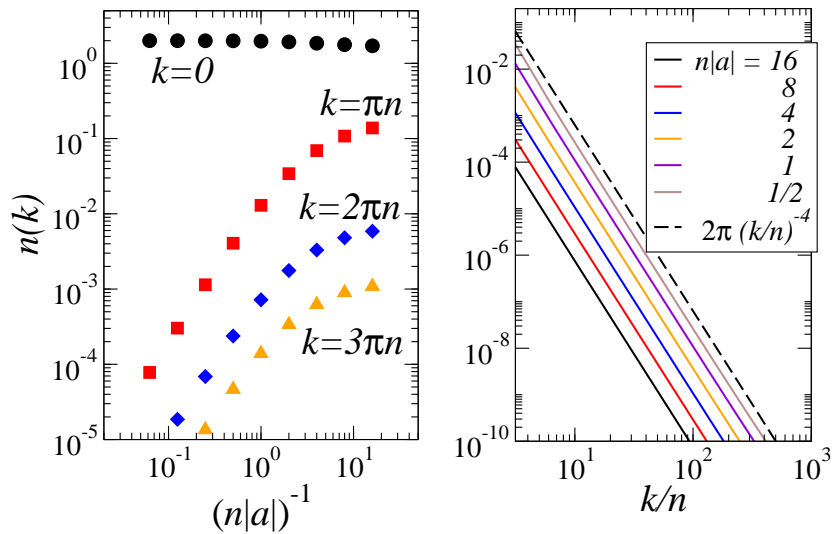}
\end{center}
\caption{\emph{Left panel:} Occupation of the lowest momentum states as a function of the 
interaction strength for the ground state of two bosons 
interacting via a $\delta$-potential. We observe that the weight lost at $k=0$ is distributed over
all momenta.
\emph{Right panel}: Momentum distribution for varying strength of the $\delta$ potential.} \label{fig:nk2bodyLL}
\end{figure}
 
 We begin our analysis of the two-body problem by investigating the Lieb-Liniger 
 two-body problem, whose results are shown in Fig.~\ref{fig:nk2bodyLL}. 
 The evolution of the momentum distribution with increasing strength of the
 potential term (decreasing $n|a|$) shows a depletion of the 
 $k=0$ contribution, and an increase of all the finite-momentum components, 
 compatible with a gradual squeezing of the two-body wavefunction $\phi(r)$
 in real space. In particular for all strengths of the interaction, the momentum
 distribution shows a characteristic decay as $1/k^4$ for all finite
 momenta -- a behavior which is consistent with that of the large-$k$ tail of $n(k)$ in the 
 $N$-body problem \cite{Olshanii03}. 
 This means that an increasing interaction strength is simply imposing a global 
 rescaling factor to the momentum distribution at finite $k$.

\begin{figure}[h!]
\begin{center}
\includegraphics[width=0.8\linewidth]{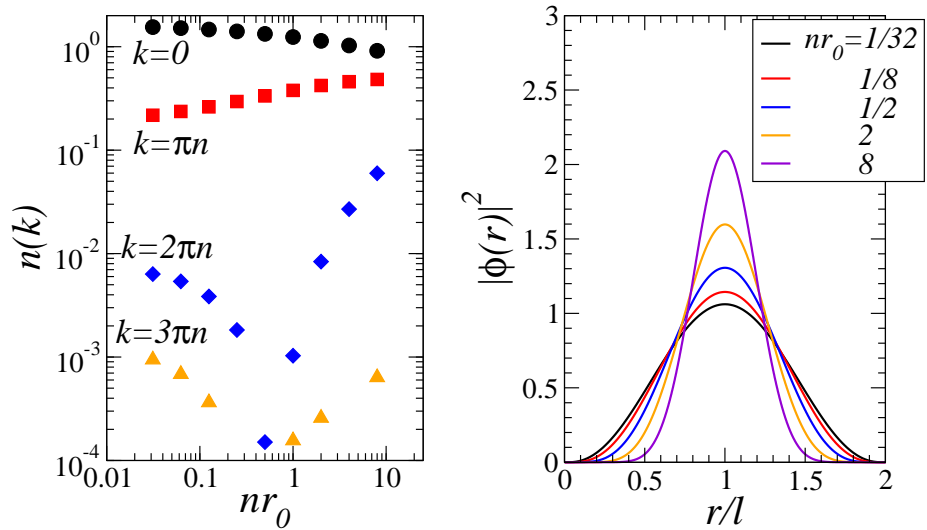}
\end{center}
\caption{\emph{Left panel:} Occupation of the lowest momentum
states for the ground state of two bosons 
interacting via a dipolar potential. We observe that, for increasing interactions,
the population lost at momentum $k=0$ is mainly transferred to momentum 
$k=\pi n$, while the population at higher momenta can be even suppressed.
This behavior is to be contrasted with that of contact interactions, Fig.~\ref{fig:nk2bodyLL}.
\emph{Right panel:} Square modulus of the two-body 
wavefunction.} \label{fig:nk2body}
\end{figure}

 The behavior in the dipolar case is radically different, as shown in Fig~\ref{fig:nk2body}. 
 First of all, in the limit $nr_0 \to 0$ the system maintains a finite occupation of 
 non-zero momentum states, due to the fact that, in this limit, dipolar bosons
 reproduce the physics of the Tonks-Girardeau gas of impenetrable particles 
 \cite{Citroetal08, AstrakharchikL08} (see Sec.~\ref{sec:nk} for further discussion). 
 Moreover, upon increasing
 the potential strength $n r_0$ we observe not only a suppression of the $k=0$
 momentum component, but also at \emph{finite} momenta $|k|\geq Q = 2\pi n$, at least for moderate
 potential strength. Hence the increase of finite momentum components, 
 associated with the squeezing of the two-body wavefunction in real space, is
 highly inhomogeneous, and it only takes place within a momentum window  
 $0 < k < Q $, while the momentum components out of this window are strongly 
 suppressed. This inhomogeneous redistribution of populations in momentum
 space, driven by increasing the interaction, results in a peculiar momentum filtering 
 effect, which is present in an even more dramatic fashion in the $N$-body problem,
 as it will be discussed in Sec.~\ref{sec:nk}. 
  

\section{\label{sec:corrfunc} One-body density matrix}

 The one-body density matrix (OBDM)
\begin{equation}
 C(x-x') = \langle \psi^{\dagger}(x) \psi(x')\rangle
 \end{equation}
 for a system of $N$ dipolar bosons in a box of length $L=N l$ with
 periodic boundary conditions, has been
 investigated via quantum Monte Carlo simulations, for 
 interaction strengths ranging between $n r_0 = 0.2$ and $n r_0 = 5$. The number 
 of particles utilized for the results shown here is $N$=30, but we performed
 calculations with $N$=15 and $N$=60 as well, in order to gauge the 
 importance of finite-size effects.        
 We generally find that simulations at a temperature 
 $k_B T \lesssim 0.1~ n r_0$ Ryd do not show any significant thermal effect, 
 and all results presented in the paper are referred to this temperature 
 range. 
 
\begin{figure}[h!]
\begin{center}
\mbox{
\includegraphics[width=0.55\linewidth]{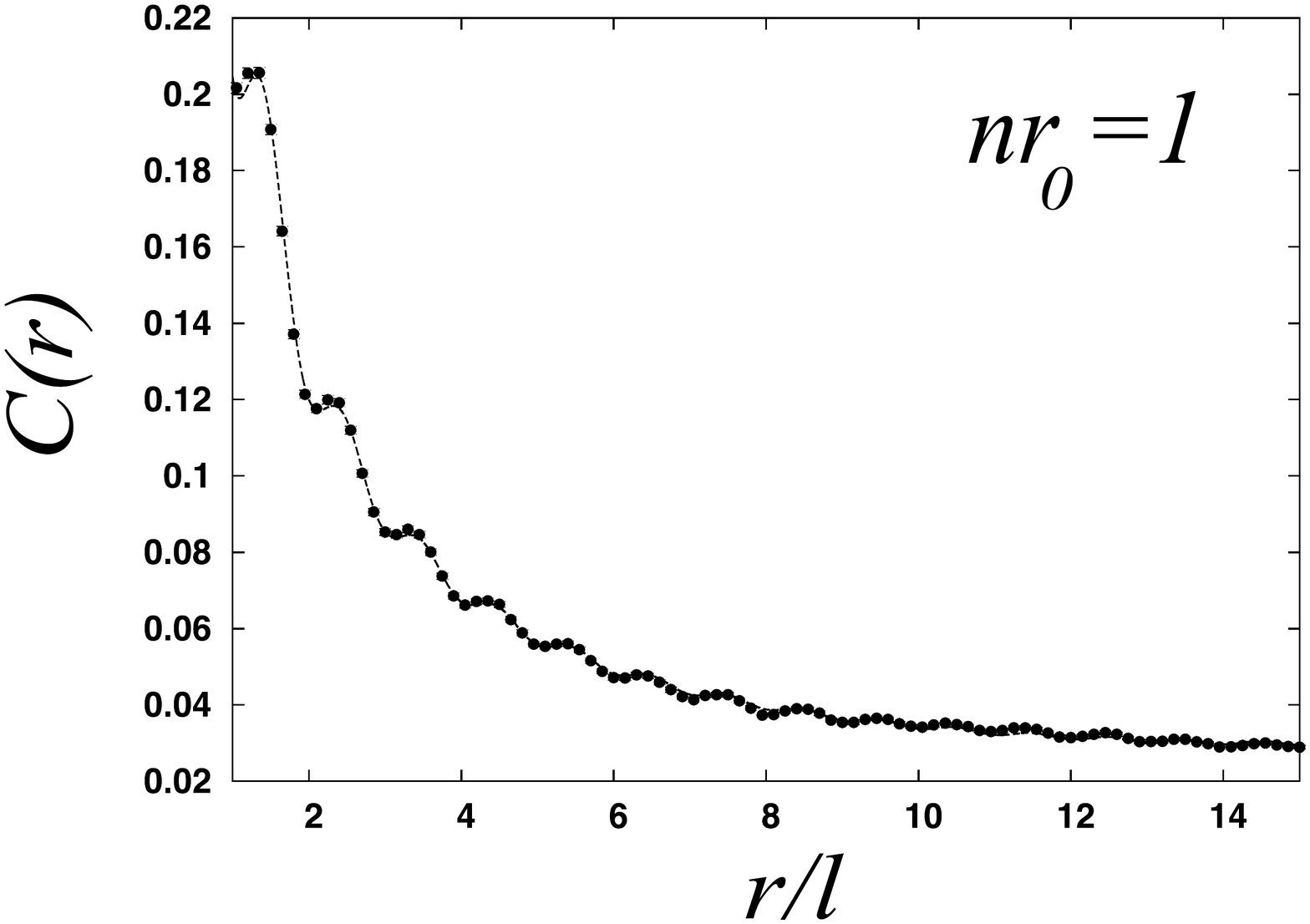}
\includegraphics[width=0.55\linewidth]{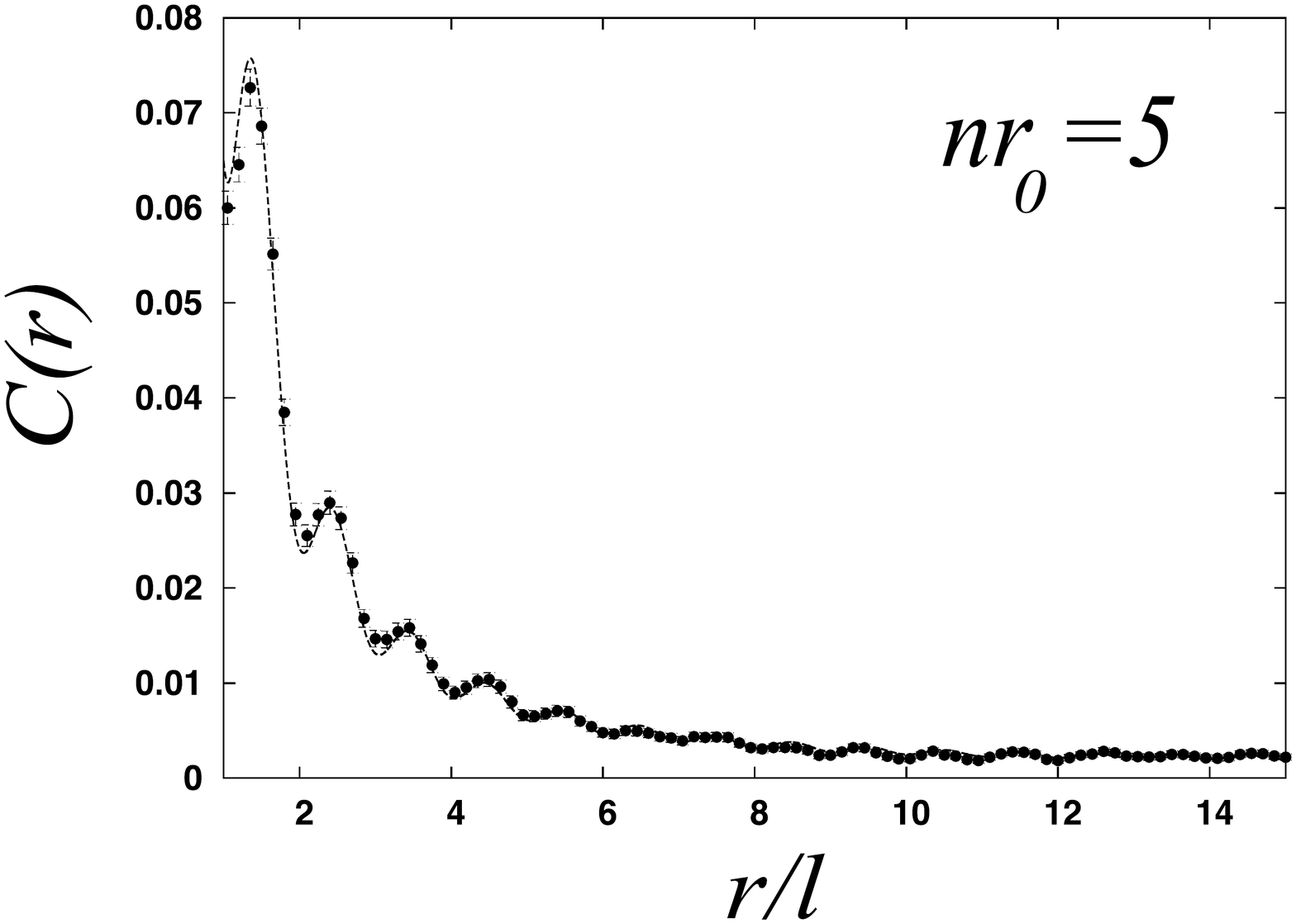}}
\end{center}
\caption{One-body density matrix for $N=30$ dipolar bosons. The dashed lines
are fits to the Luttinger liquid formula, Eq.~(\ref{e.Luttinger}), truncated to the 
$m=2$ order.} \label{fig:obdm}
\end{figure}

 Fig.~\ref{fig:obdm} shows the OBDM for two representative values of the
 potential strength $n r_0$. Strong quantum fluctuations in one dimension
 strongly suppress correlations, so that the OBDM displays an algebraic
 decay. In addition to this behavior, we observe an oscillation with period
 $1/n$, revealing the strong role played by interactions. Indeed, as we
 will see later, interactions lead to arbitrarily strong diagonal correlations, 
 which imply a crystalline structure of the particles at short range. 
 This structure is reflected in the OBDM, which exhibits a modulation 
 with an amplitude that decays spatially in the same way as the
 density correlations. 
  
   According to the general Luttinger liquid theory, density-density correlations 
   in a boson liquid display a dominant decaying behavior at large distances in the form \cite{Cazalilla04}:
   \begin{equation}
   G(x-x') = \langle \rho(x) \rho(x')\rangle \sim \frac{\cos[Q(x-x')]}{d(x-x'|L)^{2K}}
   \end{equation}
   where $K$ is the Luttinger liquid exponent; $Q = 2\pi n$ is the ordering vector of the classical Wigner crystal;
  and $d(x|L) = L |\sin(\pi x/L)|/\pi$ is the cord function.
   On the other hand, the OBDM is predicted to exhibit the following (asymptotic)
   decay
   \cite{Cazalilla04}:
   \begin{equation}
   C(r) \rightarrow_{r\to\infty} \frac{1}{d(r|L)^{1/(2K)}} \left[ b_0 + b_1 \frac{\cos(Qr)}{d(r|L)^{2K}} + \sum_{m=2}^{\infty} 
   b_m \frac{\cos(mQr)}{d(r|L)^{2m^2 K}} \right] ~.
   \label{e.Luttinger}
   \end{equation}
   Hence, besides 
  the dominant decay of the type $|r|^{-1/(2K)}$, the OBDM may also show a modulation at wavevector 
  $Q$ with an amplitude decaying as the density-density correlations, $|r|^{-2K}$, and modulations
  at higher harmonics $mQ$ with increasingly faster decay. We can successfully fit 
  our numerical results to Eq.~\ref{e.Luttinger}, and we are able to resolve harmonics up to $m=2$
  within our numerical precision. In particular, for the coefficient $b_1$ of the lowest harmonic 
  we obtain \emph{negative} values ranging in the interval $b_1\sim - 3\div5 \times 10^{-2}$. 
  The negativity of this coefficient has important consequences for the shape of the
  momentum distribution, as will be discussed in the next section. 
  
\begin{figure}[h!]
\begin{center}
\includegraphics[width=0.7\linewidth]{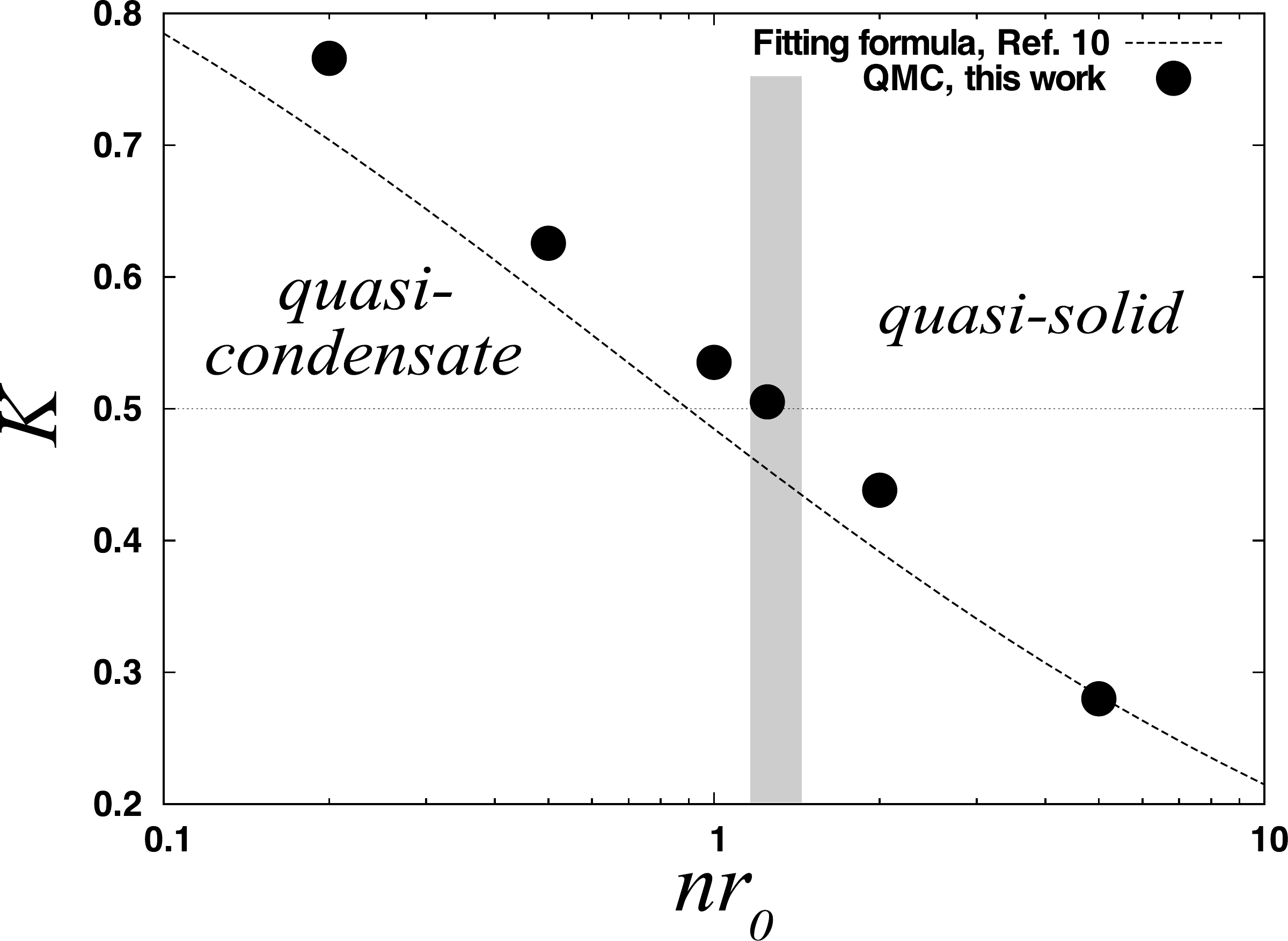}
\end{center}
\caption{Luttinger liquid exponent $K$ as a function of the interaction
strength $nr_0$. The shaded region marks the transition from 
quasi-condensation to quasi-solidity.} \label{fig:K}
\end{figure}

  In addition, fits to Eq.~\ref{e.Luttinger}
  allow to extract the exponent $K$, which is plotted in Fig.~\ref{fig:K} as a function 
  of the interaction strength $nr_0$.  $K$ has the fundamental property of governing 
  the scaling of the peak in the momentum distribution and in the static structure factor with increasing 
  system size. 
  Introducing the momentum distribution
  \begin{equation}  
  n(k) = \frac{1}{L} \int_0^{L} dr ~C(r) ~e^{ikr}
  \label{e.nk}
  \end{equation}
  and the static structure factor
  \begin{equation}  
  S(k) = \frac{1}{L} \int_0^{L} dr ~ G(r) ~e^{ikr}
  \end{equation}
  one obtains that $n(k=0) \sim L^{1-1/(2K)}$, while $S(k=Q) \sim L^{1-2K}$.
  According to the general definitions, the system exhibits \emph{quasi-condensation}
  when the peak in the momentum distribution diverges with system size ($K>1/2$);
  and it exhibits \emph{quasi-solidity} when the peak in the structure factor diverges
  ($K<1/2$). As shown in Fig.~\ref{fig:K}, increasing dipolar interactions allow
  to continuously vary the $K$ exponent from $K=1$ in the weakly interacting
  regime (characterized by fermionization \cite{Citroetal08, AstrakharchikL08}), to $K\to0$ in the 
  strongly interacting regime. In particular we find that for $nr_0 \approx 1.3$
  the Luttinger exponent crosses the value $K=1/2$, corresponding to 
  a transition from quasi-condensation to quasi-solidity. Hence dipolar interactions
  allow to continuously tune the role of correlations, driving the system from a 
  regime of dominant off-diagonal correlations to a regime of dominant diagonal 
  ones. Our estimate of the $K$ exponent stemming from the OBDM is in 
  good agreement with the semi-analytic formula given in Ref.~\cite{Citroetal08}, extracted
  from a fit to quantum Monte Carlo results for the ground state energy.

\section{\label{sec:nk} Momentum distribution}

\begin{figure}[h!]
\begin{center}
\mbox{
\includegraphics[width=0.4\linewidth,angle=270]{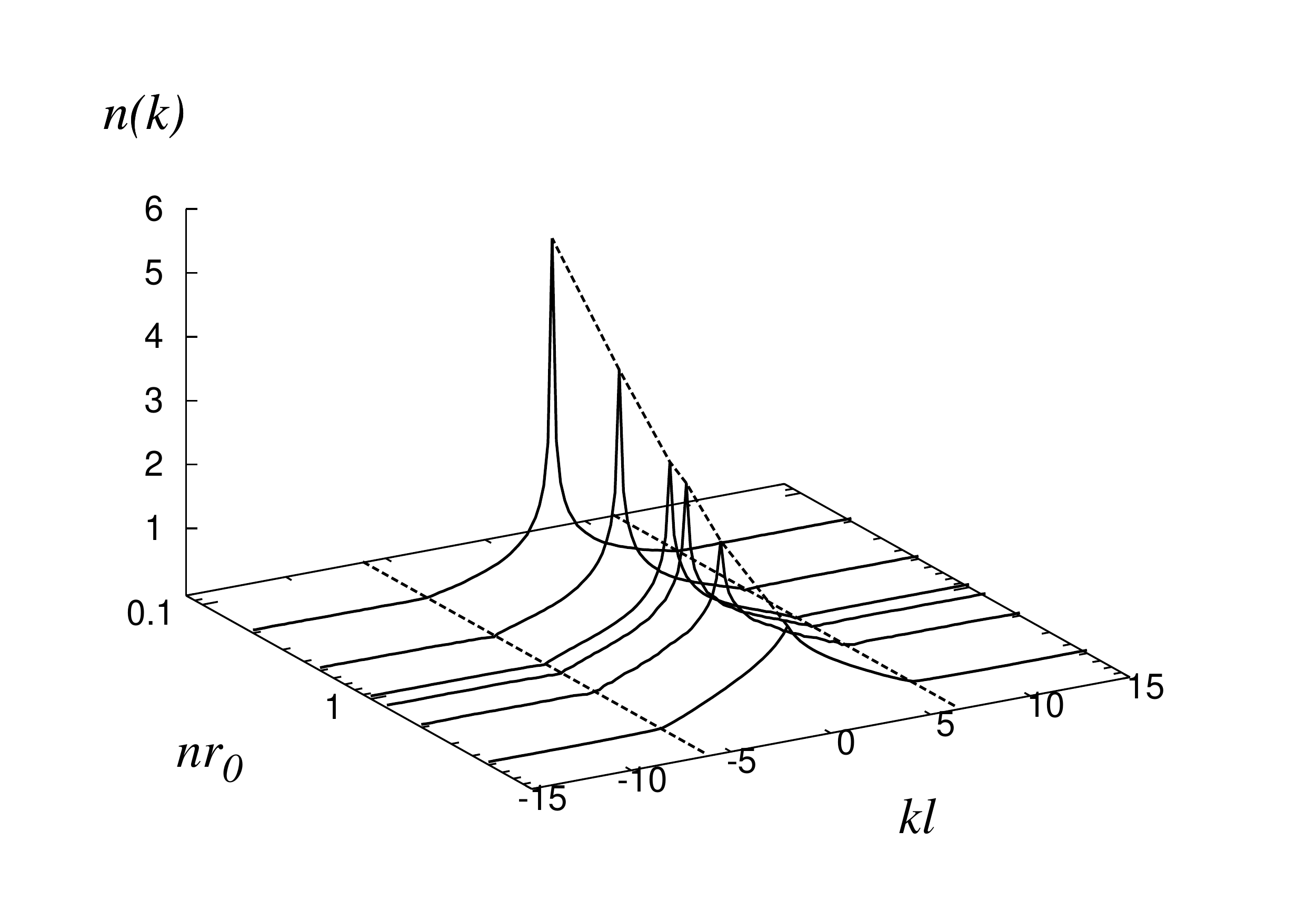}
\includegraphics[width=0.4\linewidth,angle=270]{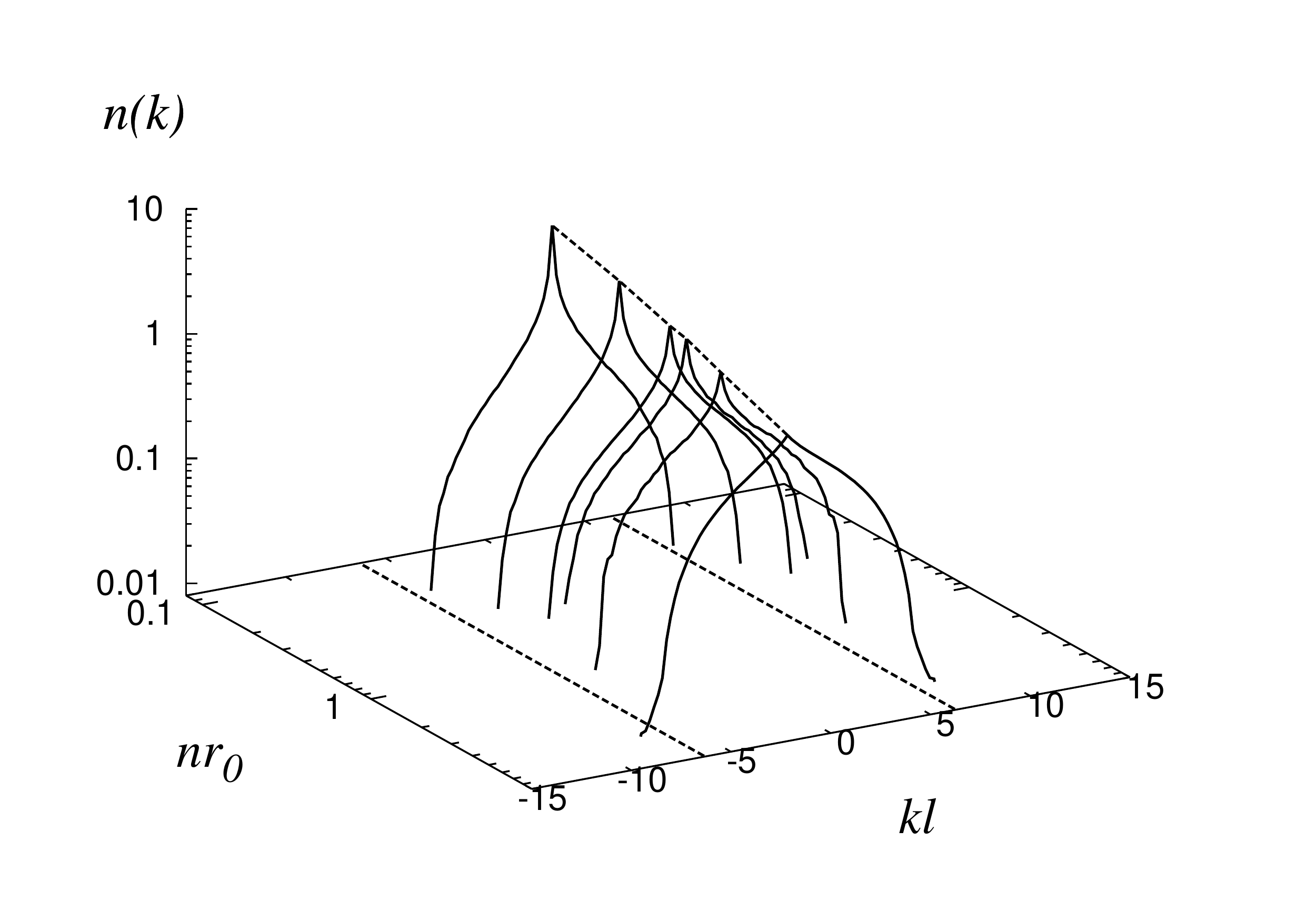}}
\end{center}
\caption{Momentum distribution as a function of the interaction strength
$nr_0$ for a system with \emph{fixed density} $n=1/l$. The dashed
lines mark the momenta $Q= \pm 2\pi n$. The right panel shows the 
distribution in logarithmic scale, to evidence the sharp suppression
for momenta $|k| > Q$.} \label{fig:nkfixn}
\end{figure}

\begin{figure}[h!]
\begin{center}
\includegraphics[width=0.5\linewidth,angle=270]{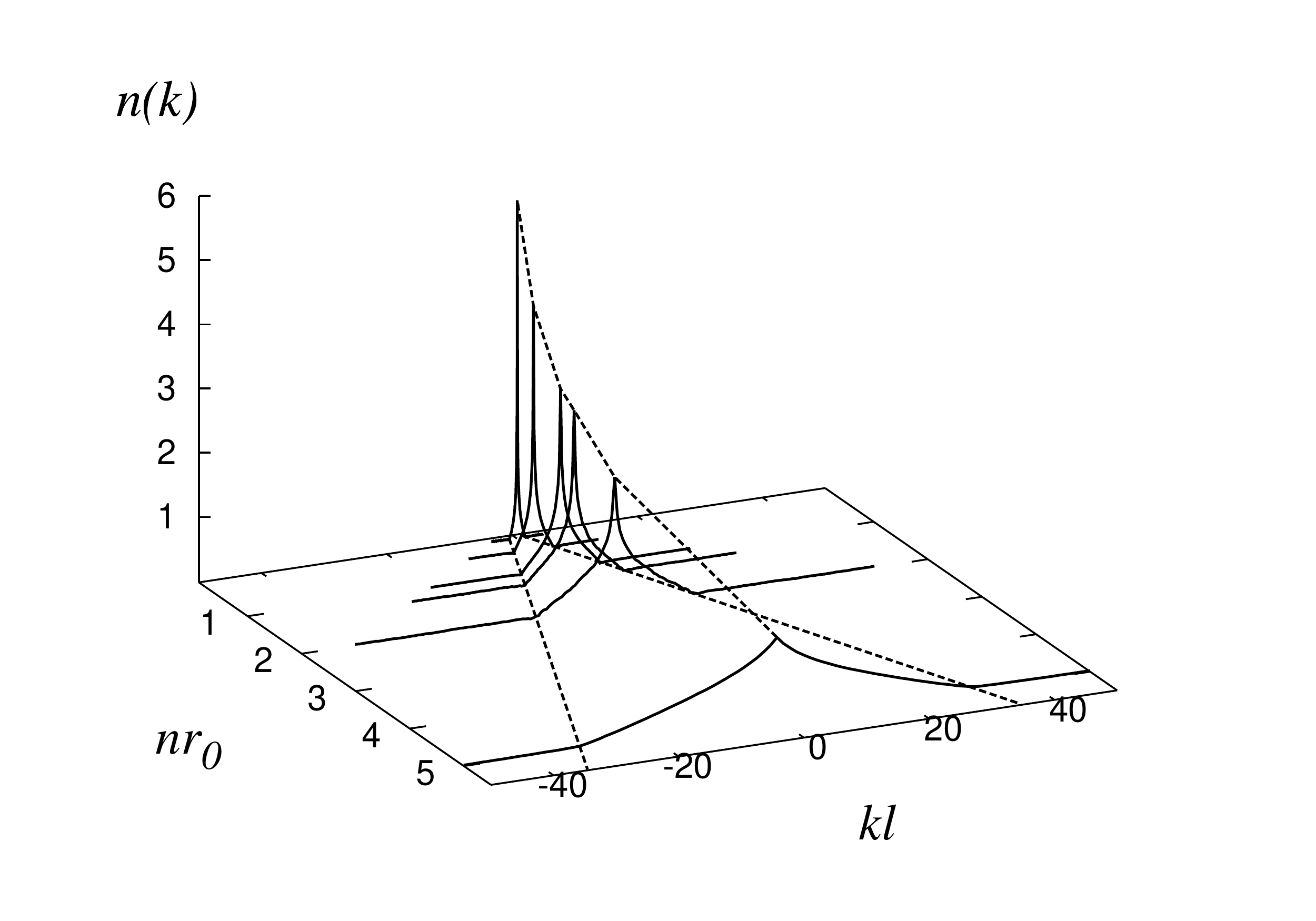}
\end{center}
\caption{Momentum distribution as a function of the interaction strength
$nr_0$ for a system with \emph{fixed dipoles}, namely a constant
$r_0$. Here the number of particles is fixed to $N=30$, and the system size 
changes when changing $nr_0$, and it goes from $L = 150 l$ for
$nr_0=0.2$ to $L=6 l$ for $nr_0=5$.} \label{fig:nkfixV}
\end{figure}

 The evolution of the momentum distribution, Eq.~\ref{e.nk}, upon changing the interaction
 strength $nr_0$ is shown in Figs.~\ref{fig:nkfixn} and \ref{fig:nkfixV}. As discussed in the 
 previous section, the OBDM exhibits a modulation term at wavevector $Q$ with a \emph{negative}
 prefactor. This means that the corresponding feature in the momentum distribution 
 is to be expected as a dip at the same wavevector. Indeed, the momentum distribution
 shows a sharp kink at $k=Q$, but it also shows a strong suppression of the
 momentum population at all momenta $|k| > Q$, observed for all investigated
values of $nr_0$. Hence we observe that the dipolar interaction
introduces a characteristic scale $Q$ in the momentum distribution, 
and a momentum filtering effect outside of the interval $[-Q, Q]$.  
This is to be contrasted with the case of contact interactions: indeed
the Lieb-Liniger model has a momentum distribution with power-law
wings, $n(k) \sim k^{-4}$ \cite{Olshanii03}, and hence it does not 
contain any special momentum scale. 
 The momentum filtering effect has the characteristic feature that, upon
 increasing the interaction strength $nr_0$, the weight lost in the $k=0$
 peak is redistributed almost exclusively over the $[-Q,Q]$ interval, similarly to 
 what we have observed in the case of the two-body problem. As a result,
 the strongly peaked distribution at low $nr_0$ evolves into an almost 
 triangular distribution for stronger $nr_0$. For extremely large values of 
 $nr_0$ one might expect that the weight lost at $k=0$ will eventually 
 start redistributing over a broader interval than $[-Q,Q]$. Nonetheless
 the persistence of Luttinger liquid behavior in the system for arbitrary
 large values of $nr_0$ leads us to conclude that the characteristic 
 periodic modulation of the OBDM, responsible for the suppression
 of the momentum distribution around $q=Q$, will also persist to much
 larger values of $nr_0$ than the ones considered here, so that the 
 momentum distribution will maintain a strong signature of dipolar
 physics around that momentum value. 
 
  Momentum filtering can be probed experimentally in two different
  setups. Working at constant density $n$, $nr_0$ can be controlled by 
  increasingly polarizing the dipoles and hence increasing $r_0$.
  This leads to the observations of Fig.~\ref{fig:nkfixn}, in which the 
  momentum distribution is always defined on the same support 
  and it changes shape from strongly peaked to triangular. A second 
  type of experiment can be carried out at fixed dipole polarization, 
  by changing the dipole density $n$, \emph{e.g.} by varying the trapping
  potential which holds the system. Fig.~\ref{fig:nkfixV} illustrates such
  a protocol, which allows to shrink arbitrarily the support
  $[-Q,Q]$ of the momentum distribution by lowering the trapping
  frequency, and hence to concentrate all the atoms to an arbitrarily
  small volume in momentum space. Yet paradoxically the system never 
  condenses; quite oppositely, as discussed in Refs. 
  \cite{Citroetal08} and \cite{AstrakharchikL08}, it \emph{fermionizes} in the limit 
  $nr_0 \to 0$, reproducing the behavior of a Tonks-Girardeau gas.

\section{Experimental realization}
 The momentum distribution is notoriously the most accessible quantity
 in trapped gas experiments, simply obtained by time-of-flight
 measurements \cite{Blochetal08}. More challenging is the realization of a quantum
 degenerate gas with dominantly dipolar interactions \cite{Lahayeetal09}. Quantum degeneracy 
 has been achieved in $^{52}$Cr \cite{Lahayeetal07, Kochetal08, Lahayeetal08}, 
 for which $s$-wave interactions
 can be suppressed by a Feshbach resonance in favor of residual magnetic dipolar
 interactions. In this system one has $V_d = \mu_0 \mu_m^2/(4\pi)$
 where $\mu_m = 6\mu_B$ (Bohr magneton) is the magnetic moment 
of $^{52}$Cr, and consequently $r_0 = 2.7 \times 10^{-9}$ m. For characteristic  
condensate densities of $10^{13}$ cm$^{-3}$ we obtain linear densities
$n\approx 2\times 10^6$ m$^{-1}$, and an effective potential strength
$nr_0 \approx 0.5 \times 10^{-2}$, which does not allow to explore the transition
from quasi-condensation to quasi-solidity described in this paper,
but it might be already sufficient to observe effects of momentum filtering.  
Polar molecules, on the other hand, can potentially feature dipolar 
interactions with characteristic lengths $r_0$ which are 3 orders of 
magnitude higher than those for $^{52}$Cr \cite{Lahayeetal09}, and they 
would hence allow to fully span the regime of interactions described 
in this paper.

Another potentially challenging issue is the confinement of dipolar
gases in one dimensional traps. Indeed confining the gas 
in an array of tubes created by a two-dimensional optical lattice
leads to \emph{quasi}-one-dimensional systems, with sizable
residual inter-tube
coupling due to the long-range nature of dipolar interactions. 
A possible solution is the use of largely spaced tubes, obtained 
via two-dimensional optical lattices created by crossing lasers
at an angle far smaller than $\pi$. Another solution is trapping
in single-mode tight atom waveguides created $e.g.$ by atom 
chips \cite{Thywissenetal99}.  

\section{Conclusions}

 In this paper we have investigated the effect of dipolar interactions
 on the off-diagonal correlations of a one-dimensional Bose gas. We have
 shown that the long-range nature of dipolar interactions changes radically
 the momentum distribution with respect to the case of contact interactions, 
 introducing a characteristic momentum scale beyond which the momentum
 distribution is strongly suppressed. The effect of momentum filtering is
 fully controlled by the gas density, and it allows to squeeze the momentum
 distribution to an arbitrarily small portion of momentum space. 
 The momentum
 distribution can also reveal the interaction-induced transition from 
 quasi-condensation to quasi-solidity, via the loss of scaling of the 
 zero-momentum peak with system size. All the phenomena discussed in 
 this paper can be observed via time-of-flight measurements on polar molecules 
 trapped in largely spaced two-dimensional optical lattices, or in tight atom waveguides.

\section{Acknowledgements}
We acknowledge very fruitful discussions with E. Orignac, and we are indebted to M. Dalmonte for pointing out a mistake 
in the comparison of our data with those of Ref.~\cite{Citroetal08}. 
This work was supported in part by the Natural Science and Engineering Research Council of Canada 
under research grant 121210893, and by the Alberta Informatics Circle of Research Excellence (iCore). MB gratefully acknowledges the support of CNRS and the 
hospitality of the Ecole Normale Sup\'erieure de Lyon during a long-term visit, in which part of this work was completed.

\end{document}